\documentstyle[prd,aps,floats,epsf]{revtex}
       \def\ssmall{\small \it}
       \def\sks{\\[2.3ex]}
       \def\skss{\\[-0.5ex]}
       \def\num{\nu_\mu}
       
       \def\rearth{R_\oplus}
       \def\s#1#2#3{#1_{\scriptscriptstyle [#2\to#3]}}

       \def\thetamix{\theta}
       \def\zen{\Theta}
       \def\asimm{{\cal A}^{\mu}_{{\scriptscriptstyle \rm U/D}}}

\begin{document}

       \onecolumn

       \title{Proposal   to  look for an  up/down asymmetry in atmospheric
       neutrinos  beyond Multi-GeV region with existing experimental data}

      \author{Shashikant R.\ Dugad,\skss
      {\ssmall Tata Institute of Fundamental Research,}\skss
      {\ssmall Colaba, Mumbai 400005, India}\sks 
       Francesco Vissani,\skss
      {\ssmall Deutsches Elektronen-Synchrotron, DESY,}\skss 
      {\ssmall Notkestra\ss{}e 85, D-22603 Hamburg, Germany;}\skss
      {\ssmall International Centre for Theoretical Physics, ICTP}\skss 
      {\ssmall Strada Costiera 11, 34100 Trieste, Italy}\sks}

       \date{March 99. IC/99/20, TIFR/HECR/99/01}
       \maketitle

       \begin{abstract}
       
       We discuss a  possible test of neutrino  oscillation hypothesis
       by  proposing  the   combined  analysis  of high   energy 
       atmospheric    neutrino  induced  muon  events  that  have been
       detected {\it around}  horizontal direction in the Kolar Gold Field
       (KGF) underground site and {\it below} the horizontal direction by
       many  large  detectors  such  as   Super-Kamiokande  and MACRO.
       Up/down asymmetry  obtained using  contained events recorded by
       detectors   at   Kamioka  site  probes  low   energy  region of
       atmospheric   neutrino  whereas,  the suggested method  probes high
       energy neutrinos. It  mainly depends  on the observations
       and it is free of  uncertainties in  neutrino flux, 
       interaction cross
       section {\it etc.} In this paper we demonstrate that the method
       is sensitive to a  region  of  oscillation parameter space
       that explains all the features  of atmospheric neutrino data in
       the  Super-Kamiokande  detector; the limiting  factor being the
       statistical strength of the KGF observations. This method
       provides  the only way  to study the  up/down  asymmetry beyond
       Multi-GeV region which is yet to be measured experimentally. 

       \end{abstract}

       \vfill

       \vskip.5truecm

       \vskip.5truecm

       \setcounter{page}{0}
       \thispagestyle{empty} \newpage

       \twocolumn
   
       \section{Motivation}\label{sec-intro}

       Evidence for oscillation of  atmospheric neutrino has been seen
       in Super-Kamiokande  detector (SK)  \cite{SK}. For a successful
       interpretation of the data, a  neutrino squared mass difference
       in the  range:
       \begin{eqnarray*}  \Delta m^2 & =  & 10^{-3} ~~ \mbox{to} 
       ~~ 10^{-2} ~~ \mbox{eV}^2 \ \ \ \ \ \ 
       \end{eqnarray*}
       and a nearly maximal mixing angle $\theta$ are suggested.
       The  interpretation in terms of  dominant $\num  \leftrightarrow
       \nu_ \tau$  oscillation  channel is favored  (even if it is not
       possible at present to exclude  the oscillations into a sterile
       neutrino).  Results  from  earlier  experiments on  atmospheric
       neutrino   anomaly  such as   Kamiokande    \cite{Kamioka}, IMB
       \cite{IMB}, Frejus \cite{Frejus}, NUSEX \cite{NUSEX} as well as
       recent    results  from   MACRO    \cite{MACRO}  and   Soudan-2
       \cite{SOUDAN2}     are    consistent  with  the   results  from
       Super-Kamiokande       detector.  The  shape  of  zenith  angle
       distribution    of   events   recorded at   Baksan   experiment
       \cite{BAKSAN}  is not in good  agreement with  the expectation.
       Also, the agreement does not  improve much by invoking neutrino
       oscillation scenario.\\

       Different  tests  of these   observations  are  essential in the
       perspective  of  passing from   ``evidence'' to  ``discovery''.
       Furthermore,  new tests should  address the  problem of precise
       evaluation  of the  parameters of  oscillation.  Long baselines
       between the  points of  neutrino production  and detection with
       controlled {\em artificial}  neutrino beams will permit crucial
       tests, especially if sufficiently large values of $L/E$ will be
       amenable.     In   this      connection,   new    results  from
       Super-Kamiokande,  MACRO and  Soudan-2  \cite{SK,MACRO,SOUDAN2}
       will  be   important  to   optimize  the   strategy  of  search
       (characteristics of the neutrino flux, design of the detectors,
       {\it etc.}) and also to interpret the result. At the same time,
       the new neutrino detectors will be able to study in more detail
       the {\em natural}  (and cost free)  atmospheric neutrinos flux.
       \\

       At    Super-Kamiokande  detector,  one  of the  most  important
       observations    is    certainly  the    up/down    asymmetry in
       $\nu_\mu$-induced muon flux. The significance of this result is
       well   beyond  the  level  that  could  be   attributed  to the
       systematic  effects  coming from  geomagnetic  effect, detector
       response {\it etc.} The most favored interpretation is in terms
       of flavour  oscillations of  neutrinos.  However, the asymmetry
       has only been  observed in  {$\num$}-induced muons with Sub-GeV
       and  Multi-GeV energy  range by  studying events  with neutrino
       interactions inside  the detector  identified as {\it fully} or
       {\it partially} contained (FC or PC) events. The events of this
       type are mostly due to low  energy neutrino interactions in the
       GeV range.\\

       It is  very   important to   extend the  search  of an  up/down
       asymmetry  beyond the  Multi-GeV  region, in  order to test the
       interpretation  in  terms of  neutrino   oscillations {\it more
       directly},  and  to  further  constrain the  allowed  parameter
       space. This can be  accomplished by using  the {$\num$}-induced
       muons in the surrounding rock. The advantage of using this data
       sample  is the  increase  in the  effective  detector  mass and
       interaction   cross  section with  the  neutrino  energy, which
       compensates for the loss in  steeply falling energy spectrum of
       neutrinos,  roughly as a  power law   (Section~\ref{sec-numu}).
       Because  of this,  the  average  energy of  neutrino  for these
       events is large compared to FC  or PC events. However, in order
       to study the up/down  asymmetry, $\asimm,$ beyond the Multi-GeV
       energy region, it is  necessary to obtain  the upward {\em and}
       downward going   $\nu_\mu$-induced muon  fluxes. In the present
       study we formulate a  proposal to achieve  this goal by mean of
       existing data.

       \section{UP/DOWN Asymmetry}\label{sec-esti}

       Many  large  detectors  like   Super-Kamiokande and  MACRO have
       recorded large numbers of  $\num$-induced muons produced in the
       surrounding    rock and   passing  (or  stopping)   through the
       detectors. However, due to  shallow depth of operation of these
       detectors, cosmic  muon flux dominates  over the $\num$-induced
       muon flux in downward  direction (except for certain directions
       in  azimuthal  angle,  where the  mountain  is  thicker and the
       shield more  efficient);  hence,  detectors  cannot distinguish
       these events from each other. The upward $\nu_\mu$-induced muon
       flux can be  measured  accurately, but this is  insufficient to
       study the up/down asymmetry mentioned above. However, detectors
       operated   at  deep    underground  Kolar   Gold  Fields  (KGF)
       \cite{compn} mines can provide  the lacking information. Due to
       large  depth of  operation  and  flat  terrain;  beyond a certain
       zenith   angle, the  flux  of  cosmic  muons is  very  small as
       compared to that of  $\nu_\mu$-induced muons in the surrounding
       rock. Being  gaseous detectors,  they cannot distinguish between 
       the particles moving in the upward and downward directions.
       Therefore, KGF  detectors measure  the sum of upward
       and downward induced muon fluxes in a given zenithal direction.
       Hence, by combining data from KGF with those from SK and MACRO,
       it is possible to  obtain the upward and  downward fluxes. Such
       an analysis permits  the study of the  up/down asymmetry beyond
       the Multi-GeV region.\\

       Some  of  the   essential   features  of  these   detectors are
       summarized in Table~\ref{tab:sites}. 
        \begin{table}[bhtp]
        \begin{tabular}{l l c c r}
  Detector & Location   & Min.\ Depth & $E^\mu_{min}$  & Ref.\ \\
           &            & (hg/cm$^2$) &  (GeV) &     \\ \hline
  Super    & Kamioka    &   2700        &   1.6         & \cite{SK}\\
Kamiokande &  Japan     &               &               &  \\ \hline
  MACRO    & Gran Sasso &   3150        &   1.0         & \cite{MACRO} \\
           &  Italy     &               &               &       \\ \hline
 Phase-1   &   KGF      &   7000        &   0.6         & \cite{KGF} \\
           &  India     &               &               &           \\
 Phase-2   &   KGF      &   6045        &   0.5         &  \cite{KGF} \\
           &  India     &               &               &           \\
        \end{tabular}        
       \caption{Features  of SK, MACRO and KGF  detectors.} 
        \label{tab:sites}
       \end{table}
       It is important to observe
       that the  energy  thresholds for  muons in these  detectors are
       different.   To get the  asymmetry  mentioned above, it is
       necessary to have the same cut on visible energy of muon in all
       the  detectors. If the  energy  threshold for KGF  detectors is
       increased,    then there  can be  a  substantial  loss of
       statistics.    Hence, it  is   necessary to   match the  energy
       threshold   of   other   detectors  with  that  of  KGF.  
       Incidentally,  this would also  lead to  further enhancement in
       the statistics of SK and MACRO detectors.\\

The  extraction of the  asymmetry parameter  requires the results
on   {$\num$}-induced  muon flux  obtained  by the    SK/MACRO and KGF
Collaborations.  At  present, these results  are available from the SK
experiment  for upward  through  going and  stopping  muons \cite{SK}.
MACRO has  published the  same for  through going muons  \cite{MACRO}.
The KGF  experiment has published  the event rate  of {$\num$}-induced
muons in rock for one part  of the existing data set; but, in order to
obtain the  flux,  informations   on the  angular  acceptance, and the
efficiencies of   trigger and detection are  also necessary. Since the
available   information  is not   sufficient to  extract the  proposed
asymmetry parameter directly from  the published experimental data, we
have evaluated the sensitivity  reach of the proposal based on certain
assumptions. Each of the assumptions mentioned below, corresponds to a
specific step in the experimental analysis.

       \begin {itemize}

       \item   {}  We   assume   that  it  is   possible  to   get the
       $\nu_\mu$-induced  muon flux at the same  muon energy threshold
       ($E_\mu > 0.5$  GeV), and the same  (zenith)  angular interval from
       all the detectors.

       \item {} SK and  MACRO detectors have  recorded a large number of
       upward going $\nu_\mu$-induced muon events that are produced in
       the  surrounding  rock.  Moreover, SK and  MACRO  being ongoing
       experiments, will collect more data. Whereas, KGF experiment has
       been stopped  in 1992 and is  estimated to  have recorded about
       250 events (Section~\ref{sec-sens}) of similar kind arriving in
       upward as well as downward  directions in a zenith angle cone of
       $55^\circ<\Theta<125^\circ$.      Hence,  statistical  error on
       downward  flux will be  relatively  much higher  as compared to
       that on upward flux. This is  due to the fact that the error on
       downward   flux,  which is   determined  by  combining  KGF and
       SK/MACRO data,  mainly depends  on the  statistical strength of
       the KGF  data. Because  of this, we  neglect the  errors on the
       experimentally  determined  {\em upward} flux,  and treat it as
       {\it true} upward  $\nu_\mu$-induced muon  flux in our estimate
       of the sensitivity.

       \item {} Phase-2 detector  at KGF site  has observed about
       23,000 cosmic muons  \cite{KGF}. Similarly, in Phase-1 detector
       also a large  number of  cosmic muons  have been  recorded. These
       observations are in excellent agreement with the predictions of
       Miyake's   empirical  relation   \cite{Miyake} up to  $\Theta <
       55^\circ$  for  Phase-1  and  $\Theta <  60^\circ$  for Phase-2
       detectors.   Beyond  zenith  angle  of   $60^\circ$ for  Phase-1
       detector ($65^\circ$  for Phase-2  detector) the rate of cosmic
       muons is negligible as  compared to the $\nu_\mu$-induced rate.
       However, in the  preceding bin,  i.e., $55^\circ <\Theta<
       60^\circ$ they are estimated to be comparable (cosmic muon rate
       is estimated to be about 30\% of total rate \cite{Pramana}). In
       order to increase the angular  acceptance, we shall assume that
       it is  possible to  determine the   $\nu_\mu$-induced muon flux
       from data in this bin, by subtracting cosmic muon flux obtained
       using Miyake's relation from observed flux in the same bin (and
       similarly    for  Phase-2    detector in  the   angular  bin of
       $60^\circ<\Theta<65^\circ$ ).

       \item {} Since the aperture area for KGF detector as a function
       of zenith  angle is  not  available, we  have  assumed it to be
       equal  for  all  bins  due to   nearly  cubic  geometry  of the
       detector.

       \item {} Systematic effects arising due to different detection
       techniques,     geomagnetic    locations,   composition  of the
       surrounding  rock {\it etc.}\ are not  considered in estimating
       the sensitivity of the proposed method.

\end {itemize}

       With  these  assumptions, the   sensitivity of the  proposal to
       oscillation parameters is obtained by the following steps:

\begin {enumerate}

       \item {} The $\nu_\mu$-induced  muon flux, ${\cal F}_\mu(\Delta
       m^2,  \thetamix, \zen_i)$, is  evaluated for each zenith
       angle    bin  of    $5^\circ$  in  the   range  of    $55^\circ
       <\zen<125^\circ$   as  a  function of  two  flavour  oscillation
       parameters (for $\nu_\mu \leftrightarrow \nu_\tau$ oscillation)
       using the  available   information on  neutrino  cross section,
       $\nu_\mu $-flux, range of muons {\it etc.} as described in the next
       section.

       \item  {}  The exposure   factor for  KGF  setup  (for  preselected
       oscillation  parameters) is obtained by  normalising total flux
       to the observed number of events as: 
       \begin{eqnarray}   
        {\cal  E} & =  &  \frac   {N_{KGF}}
        {{\sum_{i=11}^{24}}   {\cal  F}_\mu( \Delta  m^2,  \thetamix,
       \zen_i)}  \label{eqn:exp}  
       \end{eqnarray}  
       where $\zen_i = 5^\circ  \times i + 2.5^\circ.$ The 
       total number of events  observed at KGF  detectors  with the  
       suggested cuts is
       estimated to be $N_{KGF} \approx 250$ (Section~\ref{sec-sens}).

       \item {} Using this  factor, we  calculate the expected  number of events,
$N_{KGF}(\zen_i)$,  in each  angular bin for the  KGF (up+down) setup.
The  same    exposure  factor   ${\cal  E}$   is  used  to   determine
$N_{UP}(\zen_i),$  the number of upward events in the KGF data set:

       \begin{eqnarray}
         N_{KGF}(\zen_i) & = & {\cal E}\times [{\cal F}_\mu(\zen_i) +
                                  {\cal F}_\mu(180^\circ-\zen_i)],
       \nonumber\\
         N_{UP}(\zen_i) & = & {\cal E}\times {\cal F}_\mu(\zen_i)
       \label{defupdown}
       \end{eqnarray}

To obtain $N_{UP}$  experimentally, the number of SK/MACRO events have
to be rescaled to  their exposure factors,  and then multiplied by the
KGF exposure factor. However,  since the
statistics of  the SK/MACRO  detectors  are quite high  as compared to
that of  KGF,  the  estimated  number of  upward  going  events can be
treated  as a {\it  true}  number. The   sensitivity of the  method is
controlled then  by the  statistical fluctuations  of $N_{KGF}.$ Using
Eqs.~(\ref{defupdown}),  
the  total number  of  events with   exclusion of the
horizontal  bin  ($85^\circ<  \zen<95^  \circ$),  can be  obtained. We
denote them as  $N_{KGF}^{'}$ and  $N_{UP}^{'},$  for KGF and SK/MACRO
setup respectively.

       \item {} Now we can evaluate   the asymmetry between the upward
       and downward fluxes as: 
       \begin{eqnarray}
        \asimm & 
        = & \frac {N_{UP}^{'}} {N_{KGF}^{'}-N_{UP}^{'}}
        \label{eqn:asym}
       \end{eqnarray}

       \item {} Finally, the sensitivity to the oscillation parameters
       is expressed as the  significance of deviation of the asymmetry
       from unity.
        
\end {enumerate}
       
       We emphasize once again that the sensitivity has been evaluated
       only on the basis of statistical strength of the data. Possible
       systematic  effects, if any,  that could arise  while combining
       the  results from  different  detectors  have to be  taken into
       account while performing the comparison.

       \section {$\nu_\mu$-induced muon flux}\label{sec-numu}

       The estimation  is based on  the current  knowledge of neutrino
       flux, its  interaction  cross  section and muon  energy loss in
       matter. The $\nu_\mu$-induced muon flux in each angular bin can
       be written as:
       \begin{equation}
         \begin{array}{rcl}
         d{\cal F}_\mu (\Delta  m^2, \thetamix, E_\nu, \zen)&=&\\[1.5ex]
         d{\cal F}_{\nu_\mu} (\Delta  m^2,  \thetamix, E_\nu,\zen)
        &\times & \s Y\num\mu (E_\nu) 
        \label{eqn:muflux}
       \end{array}
       \end{equation}
       where we assumed that the  $\nu_\mu$-induced muon maintains the
       original neutrino direction;  an analogue formula holds for the
       ${\bar \nu}_\mu$-induced muon flux.

       The   function  $\s  Y\num\mu$  is  the  muon  yield  {\em per}
       neutrino.  The  yield  increases  with  energy of  neutrino and
       depends  on  minimum  visible  energy of  muon  required by the
       detector  ($E_{min}$).  The yield is calculated by
       considering  the  inclusive cross  section for  muon production
       $\s{d\sigma} \num\mu,$ times  the number of target nucleons per
       cm$^2$  giving rise to  sufficiently  energetic  muon above the
       threshold     energy   $E_{min}$   (0.5  GeV  in  the  present
       calculation):
       \begin{equation}
       \begin{array}{r}
        \displaystyle \s  Y\num\mu
       (E_\nu)=  \int^{E_\nu}_{E_{min}} 
       d E_\mu\,  \frac{\s{d\sigma}{\num}\mu}{d E_\mu}(E_\nu,E_\mu)
       \times\\[2ex]\displaystyle
       N_A\,  [  R(E_\mu)  -   R(E_{min}) ] \label{yield}
       \end{array}
       \end{equation}
       $N_A$  is the  Avogadro number  and $R(E_\mu)$  is the range of
       muons in the  rock. We used  neutrino flux  from \cite{BARTOL}.
       For  the  cross  section,  we  followed  the   prescriptions of
       \cite{LLS}, splitting (according to the hadronic invariant mass
       $W$) the  quasi-elastic  \cite{Llewellyn}, the  delta-resonance
       \cite{singh}   and  the   deep-inelastic  (GRV94  form  factors
       \cite{GRV}) contributions to  the charged current reaction. For
       the  range   $R(E_\mu)$ we  used  the  results   illustrated in
       \cite{LipariStanev}. \\

       For non-zero  neutrino oscillation  parameters, the $\num$ flux
       will get suppressed as:
       \begin{equation}
       \begin{array}{rcl}
       \displaystyle
        {d{\cal F}}_{\nu_\mu}(\Delta m^2, \thetamix,  E_\nu,\zen)
        &=& \\[1.5ex]
        \displaystyle 
        {d{\cal F}^0}_{\nu_\mu}(E_\nu,\zen)&\times&
        \s{P}\num\num(\Delta m^2, \thetamix, E_\nu, L_\nu)
       \label{musup}
       \end{array}
       \end{equation}
       The   survival    probability    $\s{P}\num\num$  for  $\nu_\mu
       \leftrightarrow \nu_\tau$ oscillations is given by

       \begin{equation}
       \begin{array}{l}
       \displaystyle
        \s{P}\num\num(\Delta m^2, \thetamix, E_\nu, L_\nu) =\\[1.5ex]
        \displaystyle
        1 - \sin^2(2\thetamix) \times
        \sin^2\biggl (1.27\, {\frac {\Delta m^2\,[{\rm eV}^2] 
        \cdot L_\nu \,[{\rm km}]} 
        {E_\nu \,[{\rm GeV}]}}\biggr)   \label{eqn:psur}
       \end{array}
       \end{equation}
       where $\thetamix$ is the flavour mixing angle, $\Delta m^2$ the
       square-mass  difference,  $L_\nu$ is  the  distance travelled by
       neutrino  between  its  production and  interaction  point. The
       distance travelled by  neutrino, in turn,  depends on the zenith
       angle:
       \begin{equation}
       \begin{array}{rcl}
        L_\nu & = & \sqrt{(\rearth+H_p)^2-(\rearth-H_d)^2\ \sin^2\zen}\ 
        -\\[1ex]
        &&(\rearth-H_d) \cos\zen  \label{eqn:lnu}
        \end{array}
       \end{equation}
       $\rearth$  denotes the  radius of   Earth, $H_p$  the height of
       neutrino  production  in the  atmosphere,   $H_d$  the depth of
       operation. \\

\begin{figure}[bhtp]
       \epsfxsize=3.0in
       \epsfysize=3.0in
       \epsfbox{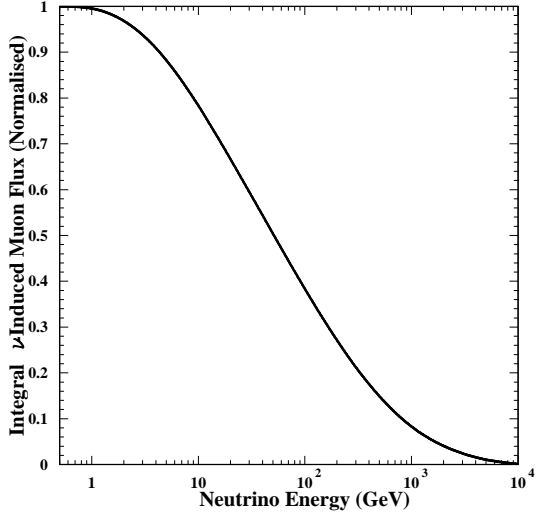}
       \caption{Parent neutrino  energy distribution for induced
       muons  with   $E_\mu>0.5$ GeV   arriving at  nearly  horizontal
       direction ($\Theta=87.5^\circ \pm 2.5^\circ$).}
       \label{fig:intfrac}
       \end{figure}

       The estimated  $\nu_\mu$-induced muon flux ${\cal F}_\mu(\Delta
       m^2,  \thetamix,  \zen_i)$ is  obtained in each  angular bin by
       integrating    Eq.~(\ref{eqn:muflux}) over  the  allowed range of
       zenith angle  
       ($\Theta_i - 2.5^\circ < \Theta  < \Theta_i + 2.5^\circ$) and
       neutrino         energy     ($E_\nu  >       E_{min}$).   
       Fig.~\ref{fig:intfrac}  shows  the normalised  integral flux of
       $\nu_\mu$-induced    muons as  a  function of   neutrino energy
       arriving in a nearly  horizontal direction  $\Theta = 87.5 ^\circ
       \pm  2.5^\circ$. As  can be  seen  from  the   figure; $(a)$
       Neutrinos   with   energies  up  to 10  GeV  are   estimated to
       contribute only to $22\%$ of  total events (the same becomes 12
       \%  at the SK  muon  energy  threshold).  In  presence of
       oscillations, which mostly  affect the lowest energy neutrinos,
       this fraction will diminish further. $(b)$ Half of the observed
       events are expected to be  originating from neutrinos of energy
       $E_\nu > 50$ GeV  (for SK muon energy  threshold, the median is
       estimated  to be at  $\sim 75$ GeV).  This  clearly illustrates
       that the  proposed asymmetry  samples more  energetic neutrinos
       than those in Sub and  Multi-GeV data sets.  
       It also indicates the necessity of
       comparing data from different detectors at the same muon energy
       threshold as stated in section~\ref{sec-esti}.

       \section  {Sensitivity of    up/down   asymmetry  to   neutrino
       oscillations}\label{sec-sens}

       The total number of  events and asymmetry  $\asimm$ in upward 
       ($95^\circ<\zen<125^\circ$)    and  downward  ($55^\circ <\zen<
       85^\circ$)       direction   are    obtained  from    estimated
       $\nu_\mu$-induced                      muon      flux     using
       Eqs.~(\ref{eqn:exp}$-$\ref{eqn:asym}).                 The
       {$\num$}-induced muon flux in  upward and downward direction is
       expected to be the  same, except for low  energy neutrinos, due
       to geomagnetic   effects\footnote{The SK  analysis of East-West
       effect proves that the theoretical expectations are well met by
       experimental data  \cite{east-west}}. Hence, the asymmetry will
       be unity for null  hypothesis, {\em  i.e.}\ if neutrinos do not
       oscillate.    Therefore, the   sensitivity to  the  oscillation
       parameters $  (\Delta m^2,  \sin^22\thetamix)$  is expressed in
       terms of significance of  deviation of asymmetry from unity, as
       a function of these parameters.\\

       The  statistical  strength of  KGF data  is the  crucial factor
       which controls the  significance of a possible deviation of the
       observable     $\asimm$  from   unity,  and    consequently the
       sensitivity to the parameters  of neutrino oscillations. The number
       of  events  recorded by  KGF  detectors in  part of  their data
       sample is 213 \cite{KGF}.  Using this and the total running time of
       Phase-2 detector  \cite{compn}, we  estimate the total number to be
       225  events.  The same  path  length  criteria  was  applied to
       Phase-1 and Phase-2  detectors, which led  to higher muon energy
       threshold  for  Phase-1  detector (Table   \ref{tab:sites}). In
       order to match the energy  threshold with that of Phase-2 detector,
       it is  necessary to reduce the  path length  cut by about 20\%,
       which in turn will  enhance the aperture,  and hence the number
       of events in the  Phase-1 detector. In  addition to this, there
       will be a further  enhancement  due to an increase in  the angular
       acceptance  by an additional  $5^\circ$.  Therefore,  in total we
       estimate   about  $N_{KGF}  \approx  250$  events from  the two
       detectors   in  KGF site   within  the  zenith  angle  range of
       $55^\circ< \zen<125^\circ$.

       The error on  up/down asymmetry is  obtained by propagating the
       statistical error on KGF observations as: 
       \begin{eqnarray}
         \Delta \asimm & = & \asimm \frac {\sqrt {N_{KGF}^{'}}}
          {N_{KGF}^{'} - N_{UP}^{'}}
         \label{eqn:signf}
       \end{eqnarray}

\begin{figure}[thbp]
       \epsfxsize=3.0in
       \epsfysize=3.0in
       \epsfbox{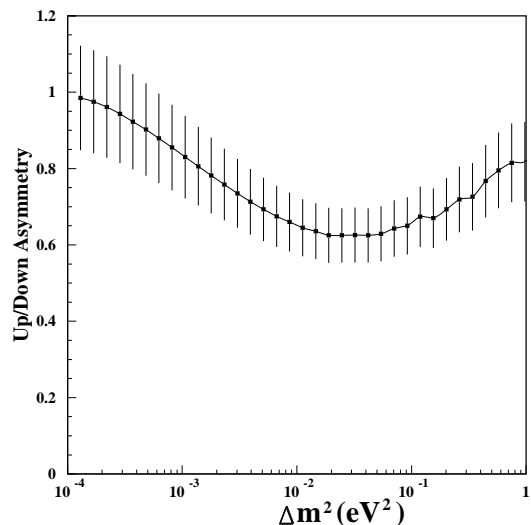}
       \caption{Up/down Asymmetry 
       beyond Multi-GeV Region,
       assuming maximal 
       mixing angle.}
       \label{fig:udasym}
       \end{figure}

       Fig.~\ref{fig:udasym} shows  the plot of the asymmetry $\asimm$
       as function $\Delta m^2$ (at maximal mixing). As it can be seen
       from the  plot, the asymmetry approaches unity at small or large
       values of $\Delta  m^2$. At small values,  neither upward going
       nor downward going  neutrinos get  oscillated significantly and
       hence   asymmetry will  be  close to  unity  due to  negligible
       oscillation. For very large values of $\Delta m^2$, oscillation
       length is quite  small as compared to the  distance travelled by
       neutrino   arriving in the  zenith  angle  range of  $55^\circ<
       \zen<125^\circ.$       Hence  the    oscillation    probability
       asymptotically    approaches to  half (at  maximal  mixing) for
       upward as well as  downward directions,  making again asymmetry
       closer  to unity.  This  implies  that the  parameter  $\asimm$
       cannot distinguish between extreme values of $\Delta m^2$. \\

       Average distance  travelled by neutrino  in upward direction in
       zenith angle region of $95^\circ < \Theta < 125^\circ$ is $\sim
       3000 ~ km$, and the median  energy of neutrino is $\sim 50$ GeV
       (Fig.\   \ref{fig:intfrac}).  Neutrino   oscillations (see Eq.\
       (\ref{eqn:psur}))   at  these  typical  values  are  maximal for
       $\Delta m^2\approx 2\cdot 10^{-2}$ eV. This value is consistent
       with the best sensitivity point

\begin{figure}[bhtp]
       \epsfxsize=3.0in
       \epsfysize=3.0in
       \epsfbox{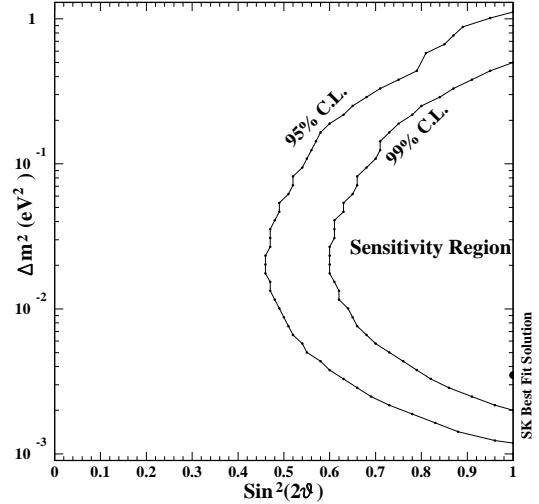}
       \caption{Sensitivity Region for 
       2-flavour oscillation}
       \label{fig:senst}
       \end{figure}

       The significance of  a deviation of  $\asimm$ from unity can be
       obtained from the estimated  value of $\asimm,$ and error on it
       (see Eqs.~(\ref{eqn:asym}$, $\ref{eqn:signf})). 
       This is translated
       to a  probability, by equating  it to the  integral of Gaussian
       probability  distribution below the  estimated significance. We
       obtain  the    sensitivity  regions  in  $(\Delta  m^2,  \sin^2
       2\thetamix)$  parameters  space, which  corresponds to a chance
       probability  of $< 5 \%$  (and $<1  \%$) to have  the estimated
       value   of  the   asymmetry  as  a  result  of  a   statistical
       fluctuation. These  regions are shown in  Fig.~\ref{fig:senst}.
       It can be seen that the proposed method is sensitive to $\Delta
       m^2$ in the  range of  $10^{-3}$ eV$^2$ to 1  eV$^2$ at maximal
       mixing.  This  more or  less   completely  spans the  region of
       oscillation parameters obtained by SK detector to explain their
       full data sample. \\

       Sensitivity of the proposed method is derived on the assumption
       that experimental data from all the detectors will be available
       at  same  visible  energy  threshold.  We show  now  that, even
       releasing  this  assumption, the  conclusions  would not change
       significantly. Using the  formalism described in Section III we
       estimated that $UP_{KGF} =  1.22\times UP_{SK}$ and $UP_{KGF} =
       1.  13\times   UP_{MACRO},$  where   $UP_{KGF},$  $UP_{SK}$ and
       $UP_{MACRO}$  denote the  upward  going  $\nu_\mu$-induced muon
       flux at  corresponding energy  threshold of  these detectors in
       the angular region  of interest. If we  are able to control the
       error on  $(UP_{KGF}/UP_{SK} -1)$ at the  level of 17\%, or the
       error on $(UP_{KGF}/UP_{MACRO}  -1)$ at the level of 28\%, then
       an  uncertainty  of $\sim  3$ \%  would be   introduced in the
       expected  number of  upward going  events  $N_{UP}'$, which was
       assumed   to  be the   {\em  true}   number of   events  in our
       sensitivity  calculations.  Since this  uncertainty is small as
       compared to statistical error  on KGF data; it would not affect
       the  sensitivity  region  significantly.  It seems  to be quite
       possible to reach the desired  control on the uncertainty. This
       is because we  only need to  get the ratio of  neutrino induced
       muon flux at  different muon energy  threshold, wherein most of
       the uncertainties on  neutrino flux,  interaction cross section
       {\em  etc.}\  tends to  get  cancelled  out  while  considering
       ratios. In this respect, it is to be noted that ratio of number
       of stopping to through-going  muons is predicted to an accuracy
       of 13 \% at SK  \cite{SK}; the  uncertainty  being dominated by
       the spectrum of primary cosmic rays.

       \section{Summary and Discussions} \label{sec-summ}

       Neutrino  oscillation signatures are  studied in SK detector by
       several methods, namely; a) ratio of electron-like to muon-like
       events  using contained  data  sample, b) up/down  asymmetry of
       contained muon events, c) shape of zenith angle distribution of
       upward going passing through muons and d) the ratio of stopping
       to   through  going  upward   muons. It  is  to be   noted that
       oscillation parameters are  better constrained by the contained
       event   data   samples,  which   corresponds  to   neutrinos of
       relatively low  energy. However, these  data may not be able to
       improve the  constraints  further,  once the  systematic errors
       start  dominating  over  statistical  errors.  Therefore, it is
       necessary to  make more  effective use of the  data sample that
       probes the high energy neutrino spectrum. 

We have shown that it is possible to obtain the up/down asymmetry
$\asimm,$  beyond  the  Multi-GeV  energy region  using the  currently
available data  from KGF, SK and  MACRO  experiments. This is a direct
measurement as it is does not  require any {\em a priori} knowledge of
neutrino  flux,  its  zenith angle   distribution,   interaction cross
section  {\em  etc.}  (the  intrinsic   asymmetry of  the  flux due to
geomagnetic  effects  only affects  the low  energy  neutrinos, which,
however,  give a  rather small  contribution  to the   neutrino parent
spectrum  shown  in  Section~\ref{sec-numu}).   
We  have  demonstrated  that this
measurement is sensitive to the allowed region of neutrino oscillation
parameter space suggested by the  recent results from SK Collaboration
\cite{SK}. We  have not  attempted an  accurate study  of systematics.
However, we checked that the inclusion of geomagnetic effects does not
lead to a significant change in the sensitivity region.

       The best fit value  obtained by  Super-Kamiokande Collaboration
       \cite{SK}  using Sub  and  Multi-GeV data  is  $\Delta m^2= 3.5
       \times  10^{-3}  ~\mbox{eV}^2,$ and  maximal  mixing; as can be
       seen from Fig.~\ref{fig:senst}, these parameters lie within the
       sensitive region of the proposed method.\\

       In  conclusion, the  proposed  asymmetry   parameter, $\asimm$,
       entails high  energy neutrinos  and it could  be obtained using
       currently   available  data  from  different   experiments. The
       analysis  of this   parameter  permits an   independent test of
       neutrino flavour conversion, having sensitivity to the range of
       neutrino oscillation  parameters  suggested by Super-Kamiokande
       experiments.

       \section{Acknowledgments} 

       This  work was   started at  the 5th  Workshop  on  High Energy
       Physics Phenomenology  (WHEPP-5) held in IUCAA, Pune, India and
       supported  by S.N.\ Bose  National  Centre for  Basic Sciences,
       Calcutta, India, by the Abdus Salam
       ICTP,  Trieste, Italy and by Tata Institute
       of Fundamental  Research, Mumbai, India.  Special thanks to W.\
       Grimus and S.\ Uma Sankar for  their interest and help provided
       in the  beginning of  this study.  S.R.D.\ would  like to thank
       B.S.\  Acharya,  V.S.\  Narasimham,  K.\  Sivaprasad  and P.R.\
       Vishwanath for many  useful discussions.  F.V.\ is grateful for
       the constant support of DESY  Theory Group, and wishes to thank
       for help S.\  Capitani, A.\  Cooper-Sarkar,  F.\ Gutbrod, R.B.\
       Hicks and  A.Yu.\  Smirnov; C.\ Arpesella for an important 
       comment; and in  particular  P.\ Lipari, who
       provided the neutrino flux  code.  We thank Alec Habig for
       correspondence.

\end{document}